# Strain-tunable magnetic and electronic properties of monolayer $CrI_3$

Zewen Wu[1], Jin Yu[2,3] and Shengjun Yuan[1,2]

1 School of Physics and Technology, Wuhan University, Wuhan 430072, China

2 Beijing Computational Science Research Center, Beijing 100094, China

3 Theory of Condensed Matter, Radboud University, Heyendaalseweg 135, 6525 AJ Nijmegen, the Netherlands

## Abstract

Two-dimensional CrI3 has attracted much attention as it is reported to be a ferromagnetic semiconductor with the Curie temperature around $45K$. By performing first-principles calculations, we find that the magnetic ground state of $CrI_3$ is variable under biaxial strain. Our theoretical investigations show that the ground state of monolayer CrI3 is ferromagnetic under compression, but becomes antiferromagnetic under tension. Particularly, the transition occurs under a feasible in-plane strain around 1.8%. Accompanied by the transition of the magnetic ground state, it undergoes a transition from magnetic-metal to half-metal to half-semiconductor to spin-relevant semiconductor when strain varies from -15% to 10%. We attribute these transitions to the variation of the *d*-orbitals of Cr atoms and the *p*-orbitals of I atoms. Generally, we report a series of magnetic and electronic phase transition in strained $CrI_3$, which will help both theoretical and experimental researchers for further understanding of the tunable electronic and magnetic properties of $CrI_3$ and their analogous.

## Introduction

Two-dimensional (2D) materials have launched great concern theoretically and experimentally because of their unique electronic and optoelectronic properties.[1-2] Typically, graphene, which is consisted of six carbon atoms in a honeycomb lattice, is a zero-gap semimetal with the carrier mobility being reported up to ~$10^5$ $cm^2V^{-1}s^{-1}$.[3] When it is chemically decorated with added atoms or cut into 1D nanoribbons, it will exhibit tunable electronic and magnetic properties.[4-6] When combined with other 2D materials to form the van der Waals (vdW) heterostructures,[7] they would exhibit much more interesting physical properties.[6, 8] Besides graphene, other 2D materials from semiconducting black phosphorus[9] to transition metal dichalcogenides[10-12] to insulating hexagonal boron nitride[13] have also been identified as important candidates for the post-silicon electronic and optical devices.[14] However, none of them is reported to exhibit intrinsic magnetism, which limits their application in spintronics.[15] Theorists predicted that most 2D materials are nonmagnetic because the thermal fluctuations under finite temperature would break the spontaneous symmetry.[16] However, a composite of monolayer $Cr_2Ge_2Te_6$ is reported to be ferromagnetic recently, which is regardless of the restriction.[17] The rising of $Cr_2Ge_2Te_6$ paves a new way in searching the long-range Ising ferromagnetism in atomically thin 2D materials, where an intrinsic magnetocrystalline could exist because of the reduction of the crystal symmetry.[18] Very recently, another ferromagnetic semiconductor chromium triiodide ($CrI_3$) appears in the research field again[19] because of its high Curie temperature in the monolayer.[20] Upon carrier doping, room-temperature magnetism is observed in $CrI_3$ due to its flat band structure.[21-23] Both experimental and theoretical research shows that monolayer $CrI_3$ is a ferromagnetic semiconductor.[18, 24] When increasing the number of the layers, ferromagnetic order is persisted within each layer, but the antiferromagnetic coupling dominates different layers.[8] Moreover, the ferromagnetic and antiferromagnetic state can be switched on and off by changing the external gate voltage.[18] When it forms heterostructures with other 2D materials like graphene, it also exhibits some topological insulating properties.[25] However, all these are concluded from a fact that $CrI_3$ is a ferromagnetic semiconductor with equilibrium lattice constant.[20]

On the other hand, strain plays an important role in determining the physical properties of 2D materials. Considering a fact that typically, $CrI_3$ is transferred on the substrate of $SiO_2$ and on other 2D materials after exfoliation,[20] the intrinsic physical properties of $CrI_3$ would be affected by the substrate due to the lattice-mismatch-induced strain.[26-27] There are some studies on the physical properties of $CrI_3$ under strain, but are based on the assumption of robust ferromagnetic ground state.[28] Meanwhile, it is challenging for experimental researchers to identify the magnetic order of the monolayer in the atomic resolution. Here, we wish to identify the magnetic ground states of $CrI_3$ under strain via *ab initio* first-principles calculations.

In this work, we present a systematic study on the tunable electronic and magnetic properties of monolayer $CrI_3$ under strain. Our results show that the Cr atoms in the unit cell are ferromagnetically aligned under compression strain, and retains the magnetic order up to a maximum tension strain around 2%, then it dramatically becomes an antiferromagnetic half-semiconductor when tension strain is further increased. During this transition, the magnetic moment on Cr atoms increases, and it undergoes a transition from magnetic-metal to half-metal to half-semiconductor, owing to the variation of the *d*-orbital of Cr atoms and *p*-orbitals of I atoms. Our results will provide a new way in understanding the magnetic ground state in monolayer $CrI_3$ and its analogous, which is useful in spintronic device designing[29] based on ferromagnetic semiconductors.[17, 30-32]

## Results and Discussions

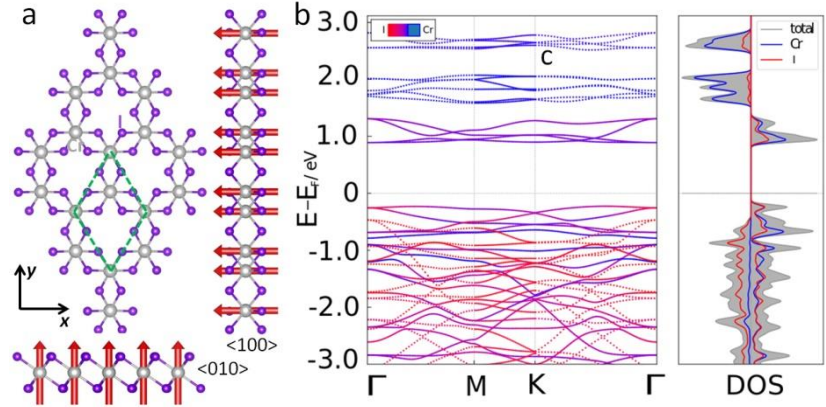

**Figure 1** Atomic and electronic structure of unstrained monolayer $CrI_3$. (a) Structural schematic of $CrI_3$ with top, front and right view. The green dashed line presents the unit cell and the red vectors indicate the spin configuration. (b) Band structure of ferromagnetic $CrI_3$. The Fermi level is set to zero. And the color indicates the attribution of Cr and I atoms. (c) Density of states of $CrI_3$ performed by spin-polarized calculations.

The atomic structure of monolayer $CrI_3$ is shown in Figure 1a, which can be simply imaged as a $\sqrt{3} * \sqrt{3}$ super cell of 1T $SnS_2$ with one point vacancy of Sn atom.[33-35] It belongs to space group C2/m containing two formula units.[36] The optimized lattice parameters are calculated to be ***a*** = ***b*** = 6.978 Å and ***c*** = 21.476 Å, which is in good agreement with the x-ray diffraction data[37] and previous DFT results[24]. Both spin-unpolarized and spin-polarized calculations are performed to get an overview of the ground state of monolayer $CrI_3$. Our result shows that the ferromagnetic (FM) state is more favorable in energy, which is 63 *meV* lower than the antiferromagnetic (AFM) state, indicating stable ferromagnetism in room-temperature. Thus, the intrinsic electronic and magnetic properties of $CrI_3$ can be represented by the FM state as shown in Figure 1b and 1c. Spin-unpolarized band structure shows that nonmagnetic $CrI_3$ is a metal with several bands crossing through the Fermi level. Upon considering the spin-polarization, the degenerated bands get split, resulting in the indirect energy gap of 1.124 *eV* and 2.169 *eV* for the spin-up (solid) and -down (dash) electrons, respectively. It is found that the spin-polarized electrons in monolayer $CrI_3$ exhibits anisotropic transport properties. For the spin-up electrons, the conduction band minimum (CBM) and valence band maximum (VBM) is located at the Gamma point and in the line from Gamma to M; for the spin-down electrons, the indirect gap originates from M and Gamma point, respectively. It is also noted that the conduction and valence band edges around the Fermi are fully spin-polarized and exclusive occupied by electrons with the same spin component, rendering a typical half-semiconductor character. Moreover, the CBM and VBM of the spin-up electrons are attributed by both Cr and I atoms, while the CBM and VBM of the spin-down electrons are attributed by Cr atoms and I atoms, respectively. Our conclusion of the intrinsic electronic structure of monolayer $CrI_3$ is further confirmed by the density of states (DOS) calculations, where sharp peaks composed of hybrid states appear around the Fermi level, suggesting strongly localized states. As a result, electrons are bounded in these states and the carrier mobility of $CrI_3$ is very slow, which can be concluded directly from those nearly flat bands around the Fermi level. Thus, in some vdW heterostructures $CrI_3$ is usually used as FM substrate to generate spin-polarized electrons.[7]

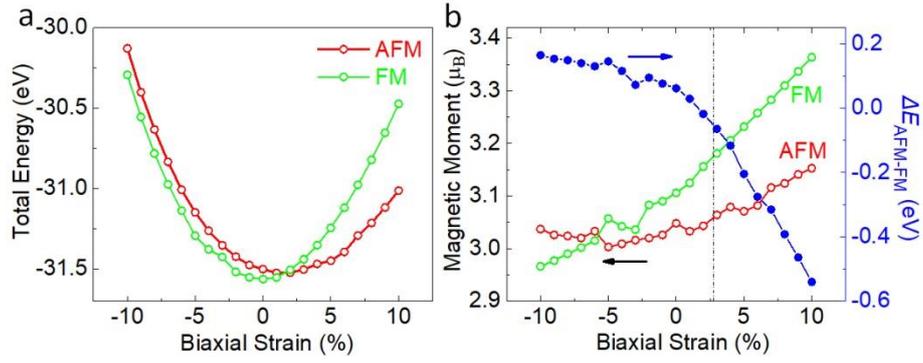

**Figure 2** Ground state of monolayer $CrI_3$ under biaxial strain. (a) Strain effect on the total energy of magnetic $CrI_3$. (b) Energy difference and magnetic moment of FM and AFM $CrI_3$ as a function of strain. Red and green symbols present the data of AFM and FM configurations, respectively.

We have shown above that pristine $CrI_3$ is a ferromagnetic semiconductor with the magnetic moment on Cr atoms being 3.106 $\mu_B$.[19,38] However, these 2D materials are usually supported by the substrate in device designing. Besides the interlayer interaction, strain induced by lattice mismatch and lattice orientation is the most common case in these 2D materials. We noted in a very recent work that when *non-collinear* spin configuration is introduced by considering the spin orbital coupling effect, monolayer $CrI_3$ will undergo a transition from FM to AFM state under compression.[39] Here, we would like to show novel electronic and magnetic properties of monolayer $CrI_3$ under biaxial strain with *collinear* spin configuration. The total energies are calculated within spin-polarized calculations for both FM and AFM configurations. It shows typically parabolic characters in Figure 2a when biaxial strain is applied. For the AFM configuration, its equilibrium lattice constant is slightly larger than that of the FM configuration. Under compression strain, the FM configuration prefers a much lower energy. To clearly show the variation of the total energy, we further plot the energy difference $\Delta E = E_{AFM} - E_{FM}$ as a function of biaxial strain in Figure 2b, where $E_{AFM}$ and $E_{FM}$ is the total free energy of the monolayer with AFM and FM configurations, respectively. Within a reasonable range from -10% to 10%, $\Delta E$ decreases monotonously and drops down to zero around 1.8%, indicating possible transition from FM to AFM. Taking the tensile strain of 3% as an example, the corresponding $\Delta E$ is 64 *meV*, which is twice larger than that calculated from the fluctuation of 300 *K*, suggesting stable AFM states at room temperature. When tension strain is further increased, the AFM configuration becomes much stable. To confirm our conclusion, we performed further calculations with DFT+U, which usually give better result for transition metals with *d* orbitals.[40-41] The parameters of *J* and *U* are chosen to be 0.7 *eV* and 2.7 *eV*, respectively, which have shown great success in predicting the magnetic anisotropic properties in monolayer $CrI_3$.[39, 42] We listed the total energies in **Table S1** of the Supporting Information. One can clearly see that though the total energy of monolayer $CrI_3$ is higher than that of the standard DFT results, the relative variation trend of $\Delta E$, $E_{AFM}$ and $E_{FM}$ are the same. As a result, the transition from FM to AFM state is obtained when the tension strain is applied.

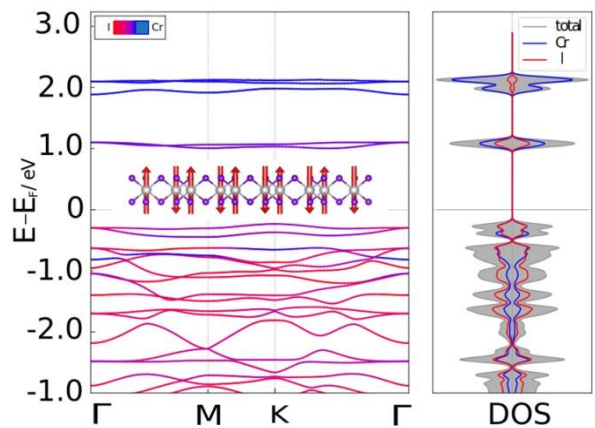

**Figure 3** Electronic structure of stretched $CrI_3$. Band structure and DOS of $CrI_3$ with tension strain of 4%. The inset shows the spin configuration

with the antiferromagnetic order. The Fermi level is set to zero.

As AFM CrI$_3$ is found to exist under tensile strain and they are less concerned, we will take tensed CrI$_3$ with a feasible value of 4% as an example to study the electronic properties. The band structure of AFM CrI$_3$ plotted in Figure 3 shows that tensed CrI$_3$ is an indirect gap spin-relevant semiconductor with the VBM and CBM at the K and M point, respectively. Different from that of FM state, spin-polarized electrons degenerate in AFM CrI$_3$ as the inversion symmetry is preserved. As a result, spin-polarized electrons are strongly localized as seen from the nearly flat bands both around 1.0 and 2.0 *eV*, which disperses with an energy window up to 0.5 *eV* in the FM configuration. The high degeneracy of the spin-polarized electrons is confirmed by the DOS as well, where the spin-up and -down electrons show identical distribution with mirror symmetry. Detailed analysis shows that though the contribution of the band structure is from the same atoms as FM CrI$_3$, the non-degenerated bands at some highly symmetric *K* points get split, indicating the symmetry broken of the $p_x$ and $p_y$ orbitals in I atoms. It is also noted that when biaxial strain is applied, the magnetic moment on the Cr atoms can be tuned by biaxial strain. When in-plane strain is applied from the compression to the tension region, the magnetic moment on Cr atoms increases for both FM and AFM configurations. Remarkably, the magnetic moment on the Cr atoms increases from 2.966 $\mu_B$ to 3.364 $\mu_B$ for FM CrI$_3$ when the strain varies from -10% to 10%. But the slope of the magnetic moment variation is much smaller for the AFM case.

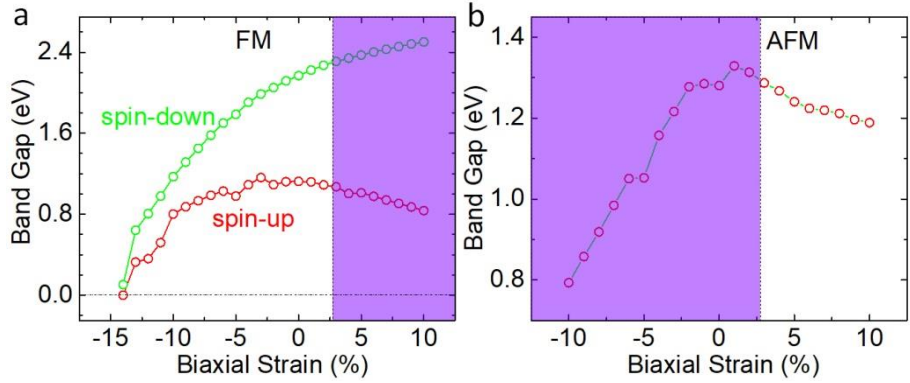

**Figure 4** Strain-dependent band gaps in $CrI_3$. (a) and (b) Spin-relevant band gaps as a function of biaxial strain. Blue and green presents spin-up and -down electrons, respectively. The shadow indicates the artificial band gap modulation in the FM or AFM configuration.

In addition to the transition and modulation of the magnetic state, the electronic properties of $CrI_3$ show an interesting response to the external strain. We show the band gap of $CrI_3$ for both spin-up and -down electrons in Figure 4. The left and right panels show the band gap modulation in the FM and AFM state, respectively. For the masked section, we will neglect the variation of the band gaps as the magnetic ground state has changed. It is obvious that the FM state is more sensitive to the biaxial strain as the slope of the modulated band gap is sharper than that of the AFM state. In the non-strained case (left panel), both spin-up and -down electrons open a gap and it shows a character of half-semiconductor (HS); when biaxial compression smaller than 13% is applied, both band gaps of the spin-up and -down electron decrease, as the band gap of spin-down electron is larger than that of the spin-up electrons, at a critical value around 14%, the spin-up gap becomes zero, while the spin-down gap remains open with 0.103 *eV*, rendering $CrI_3$ a half-metal (HM); further compressing the monolayer, both the spin-up and -down bands close, the $CrI_3$ becomes a magnetic-metal (MM). On the contrary, when tension strain is applied in a reasonable range (right panel),

AFM CrI$_3$ shows spin-relevant semiconductor (SS) characters. Both spin-up and -down electrons occupy the same band and the band gap drops down in a slow slope around 0.015 $eV$/1%. Even when CrI$_3$ is stretched by 10%, it remains open with a large band gap of 1.189 $eV$, suggesting that CrI$_3$ is a robust AFM-SS when it is under tensile strain. The transition is further confirmed by our benchmark calculations of DFT+U as shown in **Figure S1** in the Supporting Information.

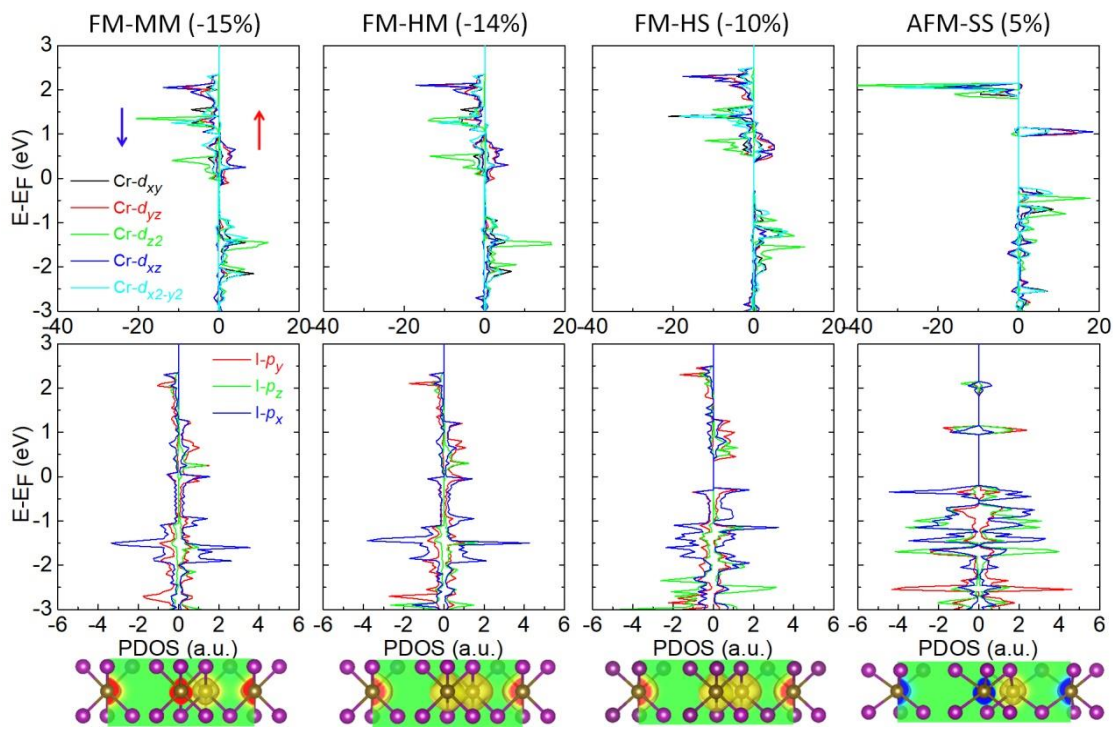

**Figure 5** Projected density of states and spin density at different strains for CrI$_3$ showing the MM-HM-HS transition. The scale bar in the slice of the spin density is set from -0.04 $e$/Å$^3$ to 0.04 $e$/Å$^3$.

To understand the mechanism of the biaxial-strain-induced electronic phase transition from MM to HM to HS to SS, we examined the projected density of states (PDOS) of CrI$_3$ under various strain as shown in Figure 5. Our result shows that the low-energy electronic properties of CrI$_3$ are mainly attributed to the in-plane components of the $d$ orbitals of Cr atoms and the $p$ orbitals of I atoms. And the $p_x$ and $p_y$ orbitals of I atoms are degenerated, which explains the degenerated VBs at some highly symmetric $K$ points in FM CrI$_3$. On the other hand, the PDOS of Cr atoms is much higher than that of I atoms above the Fermi level, suggesting that the transport properties of electrons are dominated by the variation of the $d$-orbitals of Cr atoms, while the electronic properties of holes are determined by the $p_x$ and $p_y$ orbitals of I atoms. When compression strain increases, the PDOS of all $d$-orbitals of Cr atoms shift downwards with a slightly decreasing gap for both the spin-up and spin-down electrons. However, the $p_x$ and $p_y$ orbitals of I atoms are much more sensitive to the biaxial strain, the $p$ orbitals at the conduction and valence region become hybridized, resulting in a closed gap. Specifically, at compression strain state of -15%, all the $d$ and $p$ orbitals shows peaks at the Fermi level; when compression strain decreases to -14%, only the $p_y$ orbitals of the spin-up electrons occupy the Fermi level, resulting in a MM-HM transition. For stretched CrI$_3$ with tensile strain of 10%, the peaks of the PDOS from the $p_x$ orbital of I atoms are enhanced and the conduction region is contributed by two Cr atoms with mirror symmetry. We also show the spin density of CrI$_3$ at the bottom of Figure 5. As compression strain decreases from -15% to -10%, the spin density represented by the iso-surface increases slightly, which is in good agreement with the magnetic moment modulation in Figure 2. To this end, we have shown that the electronic and magnetic properties of monolayer CrI$_3$ can be effectively tuned by biaxial strain, which is dominated by the $d_{xy}$, $d_{yz}$, $d_{z2}$ and $d_{xz}$ orbitals of Cr atoms and $p_x$ and $p_y$ orbitals of I atoms.

## Conclusions

In summary, we have systematically investigated the electronic and magnetic modulations of monolayer CrI$_3$ under biaxial strain by first-principles calculations. Applied strain yields a pronounced transition of the magnetic ground state between FM and AFM. When compression strain is applied, CrI$_3$ retains ferromagnetic. As the strain increases from -15% to 2%, a series of electronic phase transitions of MM-HM-HS-SS occur. On the contrary, it becomes antiferromagnetic under tensile strain and the band gap of AFM CrI$_3$ is robust again external strain. These modulations of electronic and magnetic properties stem from the shift of *d*-orbitals in Cr atoms and *p*-orbitals in I atoms under strain. The tunable electronic and magnetic properties in monolayer CrI$_3$ investigated in this work is helpful in understanding the magnetism in CrI$_3$ and its analogous observed by experimental researchers and would inspire extensive research interest in modulation of the electronic and magnetic properties in ferromagnetic semiconductors.

## Computational Methods

All our simulations were carried out by performing spin-polarized density functional theory (DFT) calculations as implemented in the Vienna *ab initio* Simulation Package (VASP).[43] The Perdew-Burke-Ernzerhof (PBE) pseudopotentials[44] within the general gradient approximation (GGA)[45] were used to describe the electron exchange and correlation interactions and the energy cutoff of 520 *eV* was set. The Brillouin zone was represented by a 12*12*1 mesh for the geometry optimization and the total energy calculation. And for the DOS calculation, a much denser grid of 24*24*1 was used. The atomic structure was fully relaxed with the energy convergence being $10^{-5}$ *eV*. To avoid the interlayer interaction between adjacent images, the vacuum was set to be 21 Å normal to the monolayer.

## Author information

Correspondence should be addressed to j.yu@science.ru.nl and s.yuan@whu.edu.cn .

## Acknowledgment

This work is part of the research program of the Foundation for Fundamental Research on Matter (FOM), which is part of the Netherlands Organization for Scientific Research (NWO). Yu acknowledges financial support from NSFC grant (No. U1530401) and MOST 2017YFA0303404 from the Beijing Computational Science Research Center. Yuan acknowledges financial support from the Thousand Young Talent Plan (China). Wu acknowledges the computational resource from the Supercomputing Center of Wuhan University.

Supporting Information for

# Strain-tunable magnetic and electronic properties of monolayer $CrI_3$

Zewen Wu[1], Jin Yu[2,3] and Shengjun Yuan[1,2]

1 School of Physics and Technology, Wuhan University, Wuhan 430072, China

2 Beijing Computational Science Research Center, Beijing 100094, China

3 Theory of Condensed Matter, Radboud University, Heyendaalseweg 135, 6525 AJ Nijmegen, the Netherlands

**Table S1** Total energy of monolayer $CrI_3$ with biaxial strain within DFT and DFT+U. $E_{AFM}$ and $E_{FM}$ are total energy for AFM and FM states, respectively. And $\Delta E$ is defined as $E_{AFM}$-$E_{FM}$.

|  | DFT | | | DFT+U | | |
| --- | --- | --- | --- | --- | --- | --- |
|  | -6% | 0 | 6% | -6% | 0 | 6% |
| $E_{FM}$ (eV) | -31.138 | -31.563 | -31.119 | -28.765 | -29.230 | -29.030 |
| $E_{AFM}$ (eV) | -31.008 | -31.501 | -31.395 | -28.566 | -29.147 | -29.104 |
| $\Delta E$ (eV) | 0.130 | 0.062 | -0.275 | 0.199 | 0.083 | -0.074 |

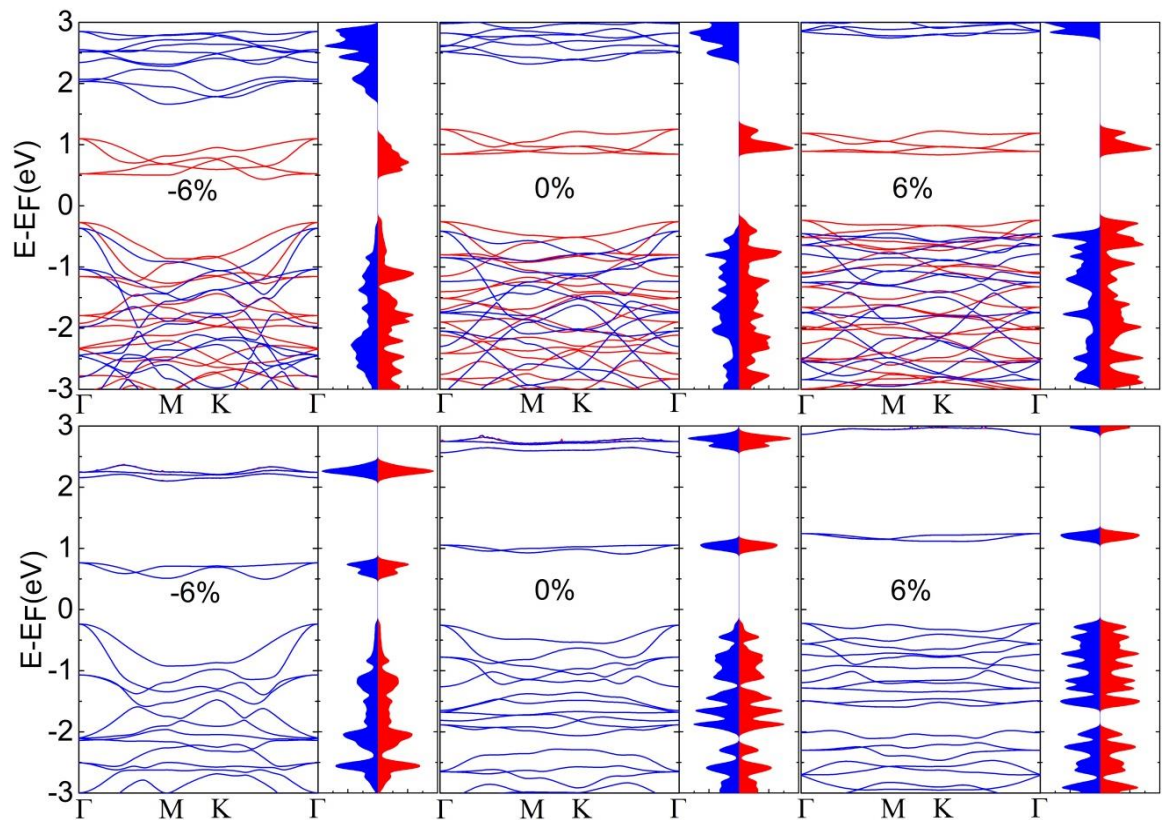

**Figure S1** Electronic structure of monolayer CrI$_3$ with biaxial strain from DFT+U calculations. Top and bottom panel represent the result for the FM and AFM state, respectively. Red and blue indicate the spin-up and spin-down electrons. The Fermi level is set to zero.